# The Foucault Pendulum Precession and the Additivity of Infinitesimal Rotations.

Lorenzo Basano
Dept. of Physics
University of Genoa - Italy

ABSTRACT – The purpose of this paper is to give an intuitive explanation of the Foucault pendulum precession (Fpp) by exploiting the easily proved result that *infinitesimal* spatial rotations about different axes in three dimensions are additive. This allows to explain in a simpler way the Foucault precession by combining physical laws, intuition and a few elementary notions of astronomy. The paper is by no means intended to replace the material given in standard texts of classical mechanics; it is only an interdisciplinary description of the Foucault precession tailored for non specialists.

## I . INTRODUCTION

It is certain that when Umberto Eco set about to write *"The Foucault's Pendulum"*, his aim was not to publish a new essay on the homonymous problem of theoretical mechanics [1]. Nevertheless, his scholarly novel acted as a scientific stimulus for people attracted by unusual physical effects.

Historically, the Foucault pendulum is important because it allowed to visualize for the first time the reality of the Earth's rotation in a room optically separated from the outside world. After that experiment, giant pendulums were built to be exhibited as popular scientific attractions in many museums scattered around the world.

The main incentive to writing this paper came from the surprising opinion expressed by a worldwide scientific celebrity, like the paleontologist Stephen Jay Gould, in the course of an interview released to the *New York* Times [2] in 1993, as reported below:

<<At the hanging pendulum, a giant steel ball suspended from a wire in the ceiling, he says: *"I've never understood why every science museum in the country feels compelled to have one of these. I still don't understand how they work, and I don't think most visitors to the museum do either."* [The interviewer tries to answer that] the pendulum is supposed to show how the suspended ball keeps swinging in a straight line while the earth rotates beneath it*,* but Dr. Gould points out that *"the pendulum itself is attached to a building that is rotating with the earth, so why should the axis of the ball not be rotating as well?">>*. Leaving aside the interviewer's imprecise and questionable statement that "the Earth rotates beneath the pendulum", I think that the doubt expressed by an outstanding scientist like Professor Gould has been unconsciously shared by entire generations of students.

In my life as a teacher, I often tried to explain the supposed "weirdness" of the Foucault precession to university freshmen (and to several friends of mine as well…) using a low level approach; but my pedagogical efforts usually turned out to be little more than a failure. Now I'd like to try an explanation that, perhaps, seems to work better than my previous attempts; for this reason I'm submitting this contribution to the wide audience of arXiv readers who pay attention not only to research but also to educational problems in physics.

In the specialized literature, the Fpp problem is usually presented following two different paths. On one side we find the standard approach adopted in University courses and in texts of theoretical mechanics [3-7] where the differential equations of motion of the pendulum are written and solved in the non-inertial local frame rigidly connected to the Earth. This approach allows to calculate both the precession frequency and the detailed trajectory of the bob. However, the majority of students, including freshmen in science and engineering, find this approach rather difficult: on one hand it requires solving a pair of coupled differential equations, which seems a bit too complex for their capabilities; on the other hand, it doesn't concede anything to physical intuition. Typically, students admit that when they write the equations of motion in the rotating Earth frame, a Coriolis term automatically appears that, after a long sequence of mathematical steps, outputs the correct precession rate. But they also add that in their opinion an intuitive explanation ought to be a different thing.

The mathematical complexity of the standard approach stimulated many authors to search for alternative procedures based on geometrically-oriented explanations; these often *link the Foucault precession to the curvature of the Earth and to parallel transport* [8-14]. As far as I know, these alternative paths are not reported in textbooks and are variously disseminated in physics journals [8-11] and in the internet [12-14]. These explanations certainly disclose the brilliant geometric visualization capabilities of the authors but also exhibit a weak point: *They are usually based on the circulation of a vector along a large closed path possibly enclosing, or touching, the Pole and regard the Earth as a very regular body*. This view hides the fact that the *origin* of the precession is *local; I'm using this term to mean that the Foucault precession need not be explained by circulating a vector around a large path*. Another drawback of these geometric methods, in my opinion, is that their pedagogical efficiency is rather low, owing to the effort required to follow the details of the explanations.

My belief is that, by focusing our attention on the *instantaneous* motion of the suspension point and on the additivity of infinitesimal rotations, the explanation of the Fpp becomes less rigorous but more intuitive, and it can be presented with success to freshmen or to twelfth grade students.

This work rests on two pillars that are discussed in the next two Sections.

**II. THE OSCILLATIONS OF A PLUMB LINE WHOSE SUSPENSION POINT IS BEING ROTATED ABOUT THE VERTICAL.**

The first pillar is a home-made experiment whose implementation is both simple and enlightening.

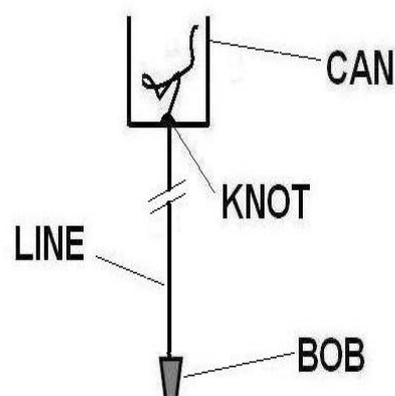

*Fig.1: Rough sketch of a home-made "apparatus" for demonstrating that the plane of oscillation of the pendulum does not follow the rotation of the can about its symmetry axis.*

Take a cylindrical can (like an emptied tomato sauce container) with a small hole bored through the center of its bottom lid; take also a plumb line

(approximately 1m long), pass its free end through the hole and make a knot just inside the can. The knot becomes the suspension point of the pendulum (Fig.1).

Now hold the can *upright*, keeping your arms extended horizontally and start the bob oscillating sideways; then begin to rotate smoothly the can around its central symmetry axis, rolling it slowly between your hand palms. You will easily note that *the direction of the plane of oscillation of the pendulum (with respect to the floor) is unaffected by the rotation of the can*. As we will see later, this simple home experiment is intimately related to the physics of the Foucault precession and clearly shows where the latter comes from. Let's now save a few remarks that will be utilized at the end: 1) The can is rotated about its symmetry axis that lies along the direction of gravity i.e. along the equilibrium position of the pendulum; 2) The angular speed with which the can is rotated is orders of magnitude greater than the diurnal angular speed of the Earth; hence, in this experiment, the Earth rotation can be reasonably neglected (in the few seconds required to rotate the can through, say, 90 degrees, the Earth rotates through an angle which is less than one hundredth of a degree). 3) (trivial but important point) Of course, in the frame of the rotating can, the plane of oscillation of the pendulum is seen to rotate with the same (but opposite) angular speed we are manually imparting to the can.

We now try to explain intuitively these facts from the point of view of an observer in the lab. *Since the symmetry axis of the can is aligned with the local gravity field, the rotation of the can does not exert any moment on the pendulum and is not dynamically coupled to it. The pendulum therefore keeps swinging in its original vertical plane*. Let's note that this conclusion does not require any calculations: symmetry considerations are sufficient to explain it. It is interesting to note that this explanation is conceptually identical to that given in texts of classical mechanics *for the special case of a pendulum oscillating at the Poles; not by chance, the Poles are the only locations on the Earth where the direction of gravity lies along the polar axis*. However the temporal evolutions of these two cases (the manually rotated can and the pendulum at the Pole), are largely different. In our home experiment the manual rotation of the can is fast enough to neglect the concurrent diurnal motion of the Earth; therefore the independence of the oscillation plane of the pendulum from the rotation of the can *is observable in a few seconds*. In the case of the pendulum oscillating at the Pole, instead, the wire is rotated only by the diurnal motion of the Earth whose angular speed is so low that *tens of minutes are needed to perceive the precessional motion*.

Now let's pose a question: would it not be possible to free us from the task of manually rotating the can and having it turned (albeit much more slowly) by the diurnal rotation of the Earth? In other words, what's wrong if we glue the can to the ceiling of the lab and wait for the Earth to rotate? (perhaps a similar reasoning might have inspired S. J. Gould's skepticism).

As it happens, it is immediate to see that our project would start off on the wrong foot. In fact, at any location different from the Poles, there is an irreducible conceptual difference between the rotation we are manually imparting to the can and the rotation produced by the Earth on the can glued to the ceiling; moreover this difference has nothing to do with the hugely different temporal scales of the two events. The essential point is that *the can of Fig.1 is constrained by us to rotate about the local vertical axis*, while *a can glued to the ceiling is constrained by the Earth to rotate around the polar axis*. Now, at any point of the Earth's surface, the angle between these two axes is equal to the complement of latitude.

Only at the Poles this angle is zero, confirming the simplicity of explaining the Foucault problem at these two special locations.

Our next goal, therefore, is *to describe how the motion of the can glued to the ceiling is mathematically related to the diurnal rotation of the Earth.* Even though, at first sight, the details of this link may not appear to be immediate, we will be helped by some basic notions usually well known to college students and astronomy lovers.

Let's consider the following analysis. The can is glued to the ceiling, which is rigidly connected to the floor which, in turn, is a portion of our horizontal plane. Therefore, *the motion of the can is the same as the motion of the horizontal plane.* It is here that elementary astronomy can enter the stage by explaining how the horizontal plane, and thus the can, is moving.

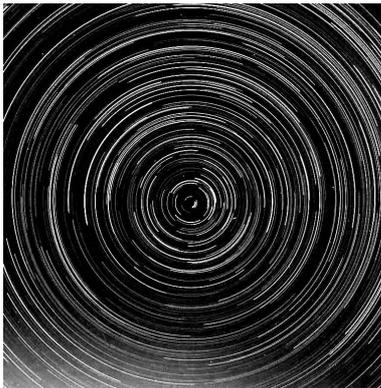

Virtually all people have run into a photograph like that shown in Fig.2. It represents a long exposure of the portion of sky centered on the North Celestial Pole, i.e. near the Pole Star.

*Fig.2  Long exposure on a starry night; the tiny bright trail near the center is created by the Pole Star.*

In our (northern) reference frame the starred sky rotates counterclokwise at an angular speed $\Omega_P = 2\pi/T_P$ where $T_P$ is the duration of the sidereal day (corresponding to a complete rotation of the Earth with respect to the "fixed stars").

Now let's consider Fig.3, which shows the basic astronomical variables judged by a hypothetical observer located at *the center C of the Earth.*

The suffixes P,V,N attached to the three $\Omega$ vectors identify, respectively, three directions: Polar, Vertical and North as clarified below.

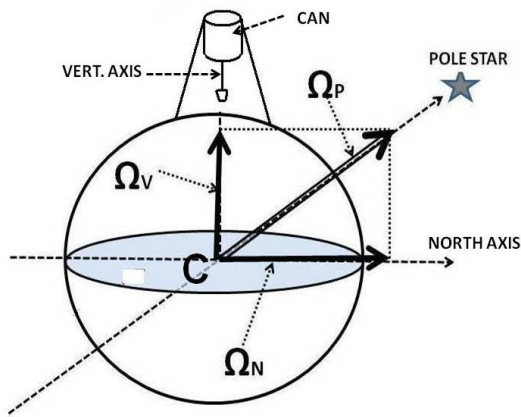

*Fig.3  $\Omega_P$ represents the angular velocity of the Earth around the Polar axis. At the top of the figure we see a rough sketch of a hugely enlarged plumb line hanging from a can fixed to the Earth*

Physically, the principal "vector" in Fig.3 is $\Omega_P$ which represents the sidereal angular velocity of the Earth around its polar axis (the reason of the quotation marks is clarified in the next Section). Provisionally, $\Omega_V$ and $\Omega_N$ are nothing else than the *geometric components* of the "vector" $\Omega_P$, respectively on the Vertical axis and on the North axis; their physical meaning is clarified below. We recall that the decomposition of a vector, in our case $\Omega_P$, along two perpendicular axes always leaves a large room for free choices. Since, in our problem, the fundamental physical direction is established by the vertical gravity field, an obvious choice is to decompose the vector $\Omega_P$ in the plane defined by the Vertical axis and by $\Omega_P$ itself, keeping $\Omega_V$ as one of the two orthogonal axes. With this choice the direction of the "North axis" in Fig.3 will be parallel to the direction of the North cardinal point in the pendulum lab.

### III. A CRUCIAL DOUBT: IS IT CORRECT TO DECOMPOSE AN ANGULAR ROTATION ALONG TWO DIFFERENT DIRECTIONS?

Here comes the second pillar of the present work.

It is well known that two-dimensional rotations (2DRs) in the x-y plane are additive in the sense that the result of two simultaneous or successive rotations $\theta_1$ and $\theta_2$ about the z-axis is equivalent to a single rotation ($\theta_1 + \theta_2$) about the same axis. A consequence of this additivity is that two 2DRs commute, i.e. the order in which they are applied is immaterial. When we shift to three-dimensional rotations (3DRs), commutativity and additivity no longer hold, as is shown in virtually all courses of classical mechanics. For example, if we rotate a box-shaped object successively around two different axes, it is manifest that the body's final orientation depends on the order of the two rotations. Now, the absence of additivity means that *3DRs cannot be handled as true vectors* (that's why rotational mechanics in space has to make recourse to more complex tools such as matrices, Euler angles or quaternions; this is also the reason for the quotation marks used above). Luckily, the good news for our problem is that, if we neglect higher order terms, *infinitesimal* 3DRs about different axes are additive and can be treated as vectors; this entails that angular velocities enjoy the same property. The demonstration is simple (it requires only an elementary knowledge of vector analysis) and is deferred to the Appendix in order to not interrupt the exposition.

At this point it is instructive to quote the following lines from Becker's text [4], pp.197-8: "….*the angular velocity is a vector which has the instantaneous direction of the infinitesimal rotation, has the magnitude of the rotation divided by the scalar dt, and obeys the parallelogram law of vector addition. If referred to a set of rectangular coordinate axes, the angular velocity* $\mathbf{\Omega}$*, like any other vector, may be expressed in terms of its components along these axes*".

Now let's go back to Fig.3 and ask: what are the *physical* meanings of $\mathbf{\Omega_V}$ and $\mathbf{\Omega_N}$ in Fig.3? According to Becker's statement, if we confine to rotations that take place in an infinitesimal time interval d$t$, the three angular rotations $d\boldsymbol{\theta_P}$, $d\boldsymbol{\theta_V}$ and $d\boldsymbol{\theta_N}$:

are related by the vector equality

$$d\boldsymbol{\theta_V} + d\boldsymbol{\theta_N} = d\boldsymbol{\theta_P} \tag{1}$$

where

$$d\boldsymbol{\theta_P} = \Omega_P \, dt \quad d\boldsymbol{\theta_V} = d\boldsymbol{\theta_P} \times sin(\varphi) \quad d\boldsymbol{\theta_N} = d\boldsymbol{\theta_P} \times cos(\varphi) \tag{2}$$

and $\varphi$ is the angle between the Polar axis and the North axis, i.e. the altitude of the Pole on the horizon: in other words the latitude of the site.

The relevance of these equations for the problem at hand is clear. If we perform an infinitesimal rotation $d\theta_V$ about the Vertical axis and (simultaneously or in succession) another infinitesimal rotation $d\theta_N$ [satisfying Eq.(2)] about the North axis, the result is an infinitesimal rotation $d\theta_P$ around the Polar axis. *For our problem it is essential that the decomposition $d\theta_P \rightarrow d\theta_V + d\theta_N$ is also valid.*

Finally, let's investigate more closely what is the separate effect on the pendulum oscillation of the component rotations $\Omega_N$ and $\Omega_V$.

I begin discussing the rotation around the horizontal North axis $\Omega_N$ (Fig.3) stating beforehand that in my opinion non-physicists will appreciate more easily an intuitive description given in the lab frame. Then, if $\Omega_N$ acted alone, at each instant the can glued to the ceiling will be subjected to an upward centrifugal acceleration $\Omega_P^2 R \cos(\varphi)$, where $R$ is the Earth radius. This acceleration acts along the vertical axis and thus it can only decrease the period of oscillation of the pendulum *without affecting the direction of the oscillation plane*. By the way, we have here a nice chance to brush up a thought-provoking problem widely discussed in texts of Physics for science and engineering [15-16].: What happens to the oscillations of a pendulum hanging from the ceiling of an elevator moving with a constant upward acceleration $a$? The answer is that $g$, the acceleration of gravity, is to be replaced by $g + a$ (this is often quoted as Einstein's first intuition on the nature of gravitation…). Coming again to our problem, it is easy to see that, at each instant, the rotation $\Omega_N$ exerts on the pendulum framework an upward pull equivalent to that imparted by the elevator's accelerated motion, causing the same decrease of the period of oscillation.

We are now close to the finish line. In fact, having established that $\Omega_N$ does not interfere with the orientation of the oscillation plane of the pendulum, we are left only with the contribution of $\Omega_V$..The latter, being a rotation about the vertical axis, produces *on the can fastened to the ceiling* a *rotation about its symmetry axis, as we did in Fig.1, but only much more slower.* In an infinitesimal interval $dt$, the oscillation plane of the pendulum lags behind the rotation of the Earth by the infinitesimal angle $d\theta_V$ which is equal to:

$$d\theta_V = \Omega_V \, dt = \Omega_P \sin(\varphi) \, dt \qquad (3)$$

Now, since *the angular speed of the Earth is constant*, it is clear that the angular lag $d\theta_V$ of the pendulum plane during an interval $dt$ is the same in all intervals of length $dt$ during a whole day and depends only on the location of our lab i.e. on the angle that the gravity field makes, at that point, with the equatorial plane. This shows that the nature of the Foucault precession is local and, therefore, need not be derived using large closed paths enclosing the Pole. The total lag $\Delta\theta$ of the plane of oscillation of the pendulum in one sidereal day can now immediately obtained by posing $dt = T_P$ in Eq.(3)., to obtain

$$\Delta\theta = 2\pi \sin(\varphi) \qquad (4)$$

which is exactly the Foucault precession formula.
.

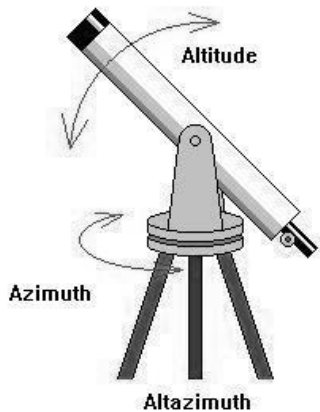

*Fig.4 A rough drawing of an altazimuth amateur telescope telescope mount.*

An interdisciplinary note may usefully complete this presentation. The vector composition of two small rotations about different axes is normally done by amateur astronomers who employ an altazimuth telescope.

As depicted by the sketchy drawing of Fig.4, the telescope can be independently rotated about the vertical axis to vary the horizontal direction (azimuth) and about the horizontal axis to vary the vertical direction (altitude).

Suppose that, at some instant, a certain star is exactly centered in the field of view of the telescope. After a small interval *dt* the star, owing to the Earth's rotation, appears to have migrated (by a very small amount) in a direction that, in general, is neither purely horizontal nor purely vertical. In order to bring the star back to the center of the field, the operator performs two very small rotations: one, about the vertical axis and one about the horizontal axis. In modern amateur telescopes a computerized system, while the Earth rotates, automatically performs this operation in real time, but the conceptual basis is the same as the one that allowed us to derive the correct Foucault precession formula. To sum up, the telescope undoes in a sequence of two very small rotations about different axes what the Earth has done in a very small rotation about the polar axis.

**SUMMARY**

It may be useful to summarize here the five-step conceptual structure of the method.

1. With reference to Fig.1, the plane of oscillation of the pendulum does not change when we rotate the can about the vertical axis. In the frame of the can, the oscillation plane is seen to precess through the same angle in the opposite sense.
2. 3-Dimensional rotations about different axes cannot be handled as vectors, but infinitesimal rotations can (and so do angular velocities).
3. By virtue of point 2 above the angular velocity $\mathbf{\Omega_p}$ of the Earth about its polar axis can be decomposed into two angular velocity *components*: one [$\Omega_p \times \cos(\varphi)$] about the horizontal North axis and one [$\Omega_p \times \sin(\varphi)$] about the vertical axis
4. The first component generates a nonessential change in the period of the pendulum without altering the direction of its oscillation plane.
5. As suggested by Fg.1 the second component accounts for the Foucault precession rate.

## APPENDIX

In this section I follow Becker's proof that infinitesimal 3DRs, and therefore angular velocities, are additive and commute. The demonstration is given in ref.[4] at pp.197-198.

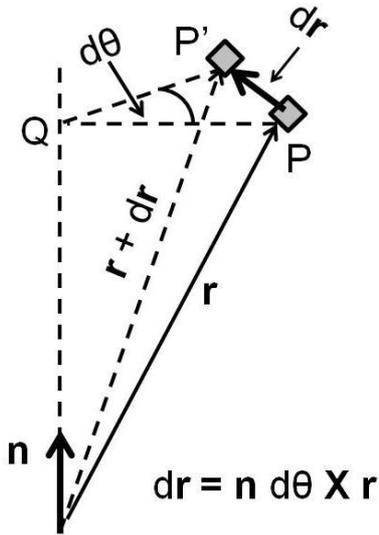

In general, the rotation of a vector ***r*** through an angle $d\theta$ about an axis whose direction in space is identified by the unit vector **n** changes **r** by:

$$d\mathbf{r} = \mathbf{n}\, d\theta \times \mathbf{r} \quad \text{(see figure)}$$

Let's now perform two successive infinitesimal rotations through angles $d\theta_1$ and $d\theta_2$ about two axes whose directions are respectively identified by unit vectors $\mathbf{n}_1$ and $\mathbf{n}_2$. First, the change of the initial vector ***r*** as a consequence of the rotation $d\theta_1$ is given by: $d\mathbf{r}_1 = \mathbf{n}_1 . d\theta_1 \times \mathbf{r}$. Now let's operate a second rotation through an angle $d\theta_2$ about an axis represented by the unit vector $\mathbf{n}_2$. The second rotation, which is not represented in the figure to avoid graphical complications, operates on the rotated vector $\mathbf{r} + d\mathbf{r}_1$ and similarly to what we did above we obtain the second vector change $d\mathbf{r}_2$:

$$d\mathbf{r}_2 = \mathbf{n}_2\, d\theta_2 \times (\mathbf{r} + d\mathbf{r}_1)$$

The total vector change ***dr*** that results from the two rotations in succession is given by the sum of $d\mathbf{r}_1$ and $d\mathbf{r}_2$:

$$d\mathbf{r} = \mathbf{n}_1\, d\theta_1 \times \mathbf{r} + \mathbf{n}_2\, d\theta_2 \times (\mathbf{r} + d\mathbf{r}_1)$$

and neglecting higher order terms:

$$d\mathbf{r} = (\mathbf{n}_1\, d\theta_1 + \mathbf{n}_2\, d\theta_2) \times \mathbf{r}$$

If the last equation is divided by the infinitesimal interval $dt$ it can be written

$$\mathbf{v} = (\mathbf{\Omega}_1 + \mathbf{\Omega}_2) \times \mathbf{r} = \mathbf{\Omega} \times \mathbf{r}$$

where the overall rotation vector $\mathbf{\Omega}$ turns out to be the vector sum of the infinitesimal "vectors" $\mathbf{\Omega}_1$ and $\mathbf{\Omega}_2$.